\documentclass[conference]{IEEEtran}
\IEEEoverridecommandlockouts
\usepackage{cite}
\usepackage{amsmath,amssymb,amsfonts}
\usepackage{graphicx}
\usepackage{xcolor}
 \usepackage{hyperref}
\usepackage{url}
\graphicspath{{figs/}}

\begin{document}


\title{Beyond the Desert Label: A Pathway Diagnostic for User-Centered Smart Mobility Service Design}
\author{\IEEEauthorblockN{Oluwasegun Adegoke}
\IEEEauthorblockA{\textit{Syracuse University} \\
Syracuse, NY, USA \\
oladegok@syr.edu}
\and
\IEEEauthorblockN{Sevgi Erdoğan}
\IEEEauthorblockA{\textit{Syracuse University} \\
Syracuse, NY, USA \\
serdogan@syr.edu}
}

\maketitle

\begin{abstract}
Smart-city mobility platforms increasingly rely on spatial screening tools to identify neighborhoods where public transit fails dependent users, but a single ``transit desert'' label can mask very different user problems: localized mismatch between service and concentrated need, or basic absence of usable service.  These call for different user-centered responses. This paper introduces a pathway-based, reproducible, data-driven diagnostic that distinguishes relative transit mismatch from minimum-service failure and reports  the specific service attributes (frequency, span, weekend service,  walking access, and  destination accessibility) driving each classification. The workflow combines open data (GTFS, ACS, LEHD, Census, and OpenStreetMap), detects spatially coherent mismatch using Local Moran's $I$, and applies an equity-informed service-failure test that centers vulnerable users. Applied to Baltimore, Philadelphia, Nashville, and Dallas, the diagnostic shows that legacy-transit cities are dominated by localized mismatch, while auto-oriented cities show broader minimum-service failure, with distinct service-deficit profiles in each case. By making the \emph{mechanism} behind an under-service label explicit, the tool supports more inclusive, user-centered smart-mobility planning across cities with different transit baselines.
\end{abstract}


\section{Introduction}
Public transit is a core service of a smart and inclusive city, yet the people who depend on it most (residents without a car, low-income households, older adults, and young people) often find it too infrequent, too slow, or too far away to meet their daily needs. Screening tools built on open urban data increasingly flag neighborhoods where transit fails these users, most commonly as \emph{transit deserts}~\cite{jiao2013}. Such tools are widely used for equity screening and mobility planning~\cite{aman2020,guo2024,jiao2017}. However, the same transit desert label can describe very different user problems. In one neighborhood, transit-dependent residents may face a localized mismatch between where service runs and where need concentrates. In another, they may face a broad absence of minimum usable service, with buses too infrequent, no weekend trips, or stops beyond walking distance. Each implies a different user-centered response, yet a binary label cannot distinguish them.

This distinction is central to designing user-centered smart-mobility services, but it remains difficult to operationalize across cities with different transit baselines. Relative, city-normalized gap methods are useful for identifying within-city disparities but can obscure whether the overall level of service is high or low~\cite{jiao2017,jiao2013}. Threshold-based approaches identify minimum-service failures but may over-label low-density areas unless interpreted alongside measures of need~\cite{karner2024}.  For user-centered mobility, where the goal is to understand the lived experience of specific rider groups, this ambiguity is not a technicality: it determines whether the right intervention is service realignment, better pedestrian access, or new baseline coverage. In prior work, we developed a supply--demand gap framework 
using LISA clustering to identify transit deserts across four 
U.S. cities \cite{adegoke2026}, but that framework assigns a 
single label without distinguishing \textit{why} a tract is 
classified.

We develop a reproducible, data-driven diagnostic that reveals why a neighborhood is classified as under-served, so that planners and mobility-service designers can identify which problem to address. The workflow combines open data (GTFS schedules, ACS demographics, LEHD employment, Census tract boundaries, and OpenStreetMap-derived built-environment measures) to construct tract-level supply and user-vulnerability features. A relative pathway identifies where transit supply is low relative to local user need and forms spatially coherent clusters (Local Moran's $I$); an equity-informed absolute pathway identifies tracts that fail minimum service criteria and also exhibit above-median latent transit vulnerability, centering vulnerable users rather than low-density geography alone. Crucially, the diagnostic also reports which service attributes (frequency, service span, weekend service, walking access, and destination accessibility) are most deficient, translating an abstract label into the concrete dimensions of a rider's experience. We apply the diagnostic to Baltimore, Philadelphia, Nashville, and Dallas. 

This paper makes three contributions:
\begin{itemize}
\item We move beyond a single ``transit desert'' label by 
distinguishing \emph{which} user problem a neighborhood 
faces: a localized mismatch between service and need, or 
a basic absence of usable service.
\item We combine gap analysis, spatial clustering, and 
threshold-based classification into a reproducible open-data 
workflow for user-centered mobility screening.
\item We show that identifying \emph{how} a tract is 
classified, and reporting its specific service deficits, 
makes transit desert classifications more interpretable 
and comparable across cities with different transit baselines.
\end{itemize}

\section{Related Work}
\subsection{Transit deserts, gaps, and user need}
Transit desert analysis frames under-service as a spatial gap between transit supply and transit need ~\cite{jiao2013, jiao2017, aman2020}. Supply--demand gap methods have been extended with richer vulnerability indicators and multidimensional supply measures including frequency, connectivity, and destination accessibility \cite{guo2024, welch2013}. However, most methods produce a single binary label without distinguishing whether the underlying problem is relative mismatch or minimum-service failure, a distinction that Karner et al. \cite{karner2024} identify as central to valid cross-city equity comparison.  Accessibility research distinguishes the \textit{distribution} of accessibility from the achievement of \textit{minimum standards} \cite{geurs2004}. From a user-centered perspective these are distinct experiences: being worse-off than one's neighbors is not the same as lacking usable service outright. Smart-mobility research increasingly emphasizes this gap between system-level metrics and the rider's lived experience \cite{jittrapirom2017}. Recent work by Yadav et al. \cite{yadav2024} distinguishes transit desert zones from transit-stressed zones, and multi-criteria frameworks have been developed to determine service-zones where demand-responsive services such as microtransit would be more beneficial \cite{erdogan2024}. We build on this research by classifying tracts according to \textit{which} pathway identifies them and reporting user-facing deficit profiles behind each label.

\section{Data and Study Areas}
 We apply the diagnostic to four U.S. metropolitan areas: Baltimore and Philadelphia (legacy transit systems with dense fixed-route networks) and Nashville and Dallas (auto-oriented Sunbelt systems with more limited service). We use the same four cities from our earlier work \cite{adegoke2026} to test 
whether the two-pathway approach reveals different types of under-service across contrasting transit contexts. All measures are computed at the Census tract level. The workflow uses publicly available data: GTFS static feeds for service characteristics, ACS 5-Year Estimates for vulnerability indicators, LEHD LODES for employment accessibility, Census TIGER/Line for tract boundaries, and OpenStreetMap for built-environment features. Tracts with zero population or consisting primarily of water or parkland are excluded. Full data-source details and processing steps are described  in \cite{adegoke2026}.

\section{Diagnostic Workflow}
\subsection{Design Goals}
\begin{figure}[t]
\centering
\includegraphics[width=\columnwidth]{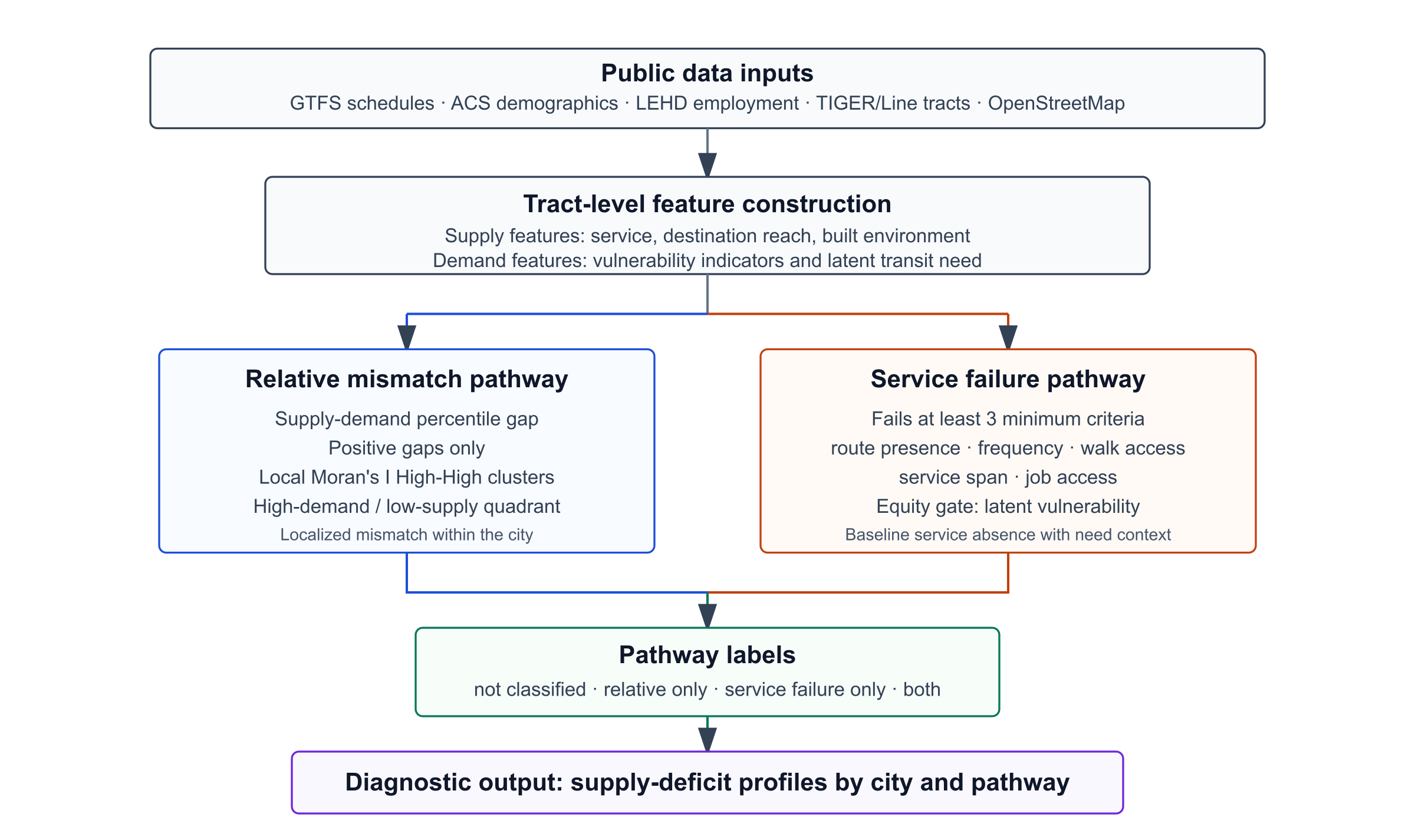}
\caption{Pathway-based diagnostic workflow for distinguishing relative transit mismatch from equity-relevant service failure.}
\label{fig:workflow}
\end{figure}

The workflow classifies the \emph{mechanism} behind a transit desert label rather than assigning only a binary under-served/not-under-served outcome. Three goals guide it. First, it should preserve relative spatial information, because local mismatch between supply and user need is central to transit desert analysis. Second, it should retain absolute service information, because a tract in a low-baseline system may rank favorably within its city while still failing basic service conditions that users experience. Third, it should incorporate equity context into the absolute pathway, so that low service alone does not automatically classify low-density or low-need areas as deserts. These goals lead to a pathway-based classification (Fig.~\ref{fig:workflow}).

\subsection{Feature Construction}
The workflow constructs tract-level supply and user-need feature sets from public data. The supply set captures three dimensions of effective transit access: scheduled service (stop connectivity, route coverage, frequency, span, weekend service), destination accessibility (walking access to transit, job accessibility, access to essential points of interest), and built-environment support (population density, intersection density, land-use mix, sidewalk infrastructure). Features are standardized within each study area and combined into a composite supply score that orders tracts by relative transit access within each city; to reduce redundancy among correlated variables, feature weights are adjusted within each group using a correlation-penalty approach adapted from composite-indicator methods~\cite{rathod2024}. This penalty down-weights variables that are strongly correlated with other indicators in the same supply dimension, reducing double-counting while preserving indicators that contribute distinct information.

The user-need set represents transit vulnerability rather than observed ridership. It combines ACS indicators of zero-vehicle households, poverty, minority population, elderly population, and youth population. Because vehicle ownership can understate need in auto-oriented regions, we also construct a latent-vulnerability variant that partially replaces  observed zero-vehicle households with a forced-car-ownership proxy, which flags low-income households whose vehicle ownership likely reflects the absence of viable alternatives rather than preference. This variant is used only as the equity information for the service-failure pathway.

\subsection{Relative Mismatch Pathway}
The relative pathway identifies tracts where user need is high relative to supply within the same urban context. We compute a percentile-based supply--demand gap:
\begin{equation}
Gap_i = PctRank(Demand_i) - PctRank(Supply_i),
\end{equation}
where higher values indicate a tract ranks higher in vulnerability than in supply. Percentile ranks are used because supply and demand may have different distributions and because the pathway targets relative mismatch within each city rather than raw cross-city magnitudes. Only positive gaps are retained. We then compute Local Moran's $I$ with Queen contiguity weights to identify spatially coherent high-gap clusters~\cite{anselin1995}. A tract is assigned to the relative pathway if it lies in a statistically significant High--High cluster or falls in the high-demand/low-supply quadrant defined by city medians.

\subsection{Service Failure Pathway}
The service-failure pathway identifies tracts that fall below minimum service conditions. A tract fails the service screen if it fails at least three of five criteria: route presence within walking distance; peak headway of 30 minutes or less; at least 25\% of tract area within walking distance of a stop; at least 12 hours of daily service span; and at least 5{,}000 jobs reachable within 45 minutes by transit. The 30-minute peak-headway and 12-hour service-span thresholds follow transit quality-of-service guidance that treats these as lower bounds for usable fixed-route service \cite{tcqsm2013}. The 45-minute job-access window follows cumulative-opportunity practice, which counts the opportunities reachable within a fixed travel-time budget \cite{higgins2022accessibility}. The 5{,}000-job cutoff is a conservative opportunity floor, informed by employment-center literature that treats roughly 5{,}000 jobs as a lower-bound meaningful employment concentration \cite{tcrp16urbanform1996,shearmur2007employment}; sensitivity to these cutoffs is examined in Section VI.   Because low service alone does not imply unmet need, the pathway includes an equity condition.

\subsection{Diagnostic Outputs}
Each tract receives one of four labels: not classified, relative mismatch only, service failure only, or both. These make the mechanism explicit: a relative-only tract suggests localized mismatch within an otherwise functioning service context; a service-failure-only tract suggests minimum-service absence in a vulnerable area; a both-pathways tract indicates the two overlap. The workflow also produces \emph{supply-deficit profiles}: for each classified tract we identify the lowest-scoring supply features and summarize, by city, which deficits appear most often. These outputs indicate whether classified areas are associated primarily with frequency, service span, walking access, destination accessibility, or built-environment limitations.

\section{Results}
We organize results around the classification behavior of the workflow rather than individual case studies, evaluating whether pathway labels clarify different mechanisms of user-facing under-service across contrasting contexts.

\subsection{Pathway Composition Across Cities}
\begin{figure}[t]
\centering
\includegraphics[width=\columnwidth]{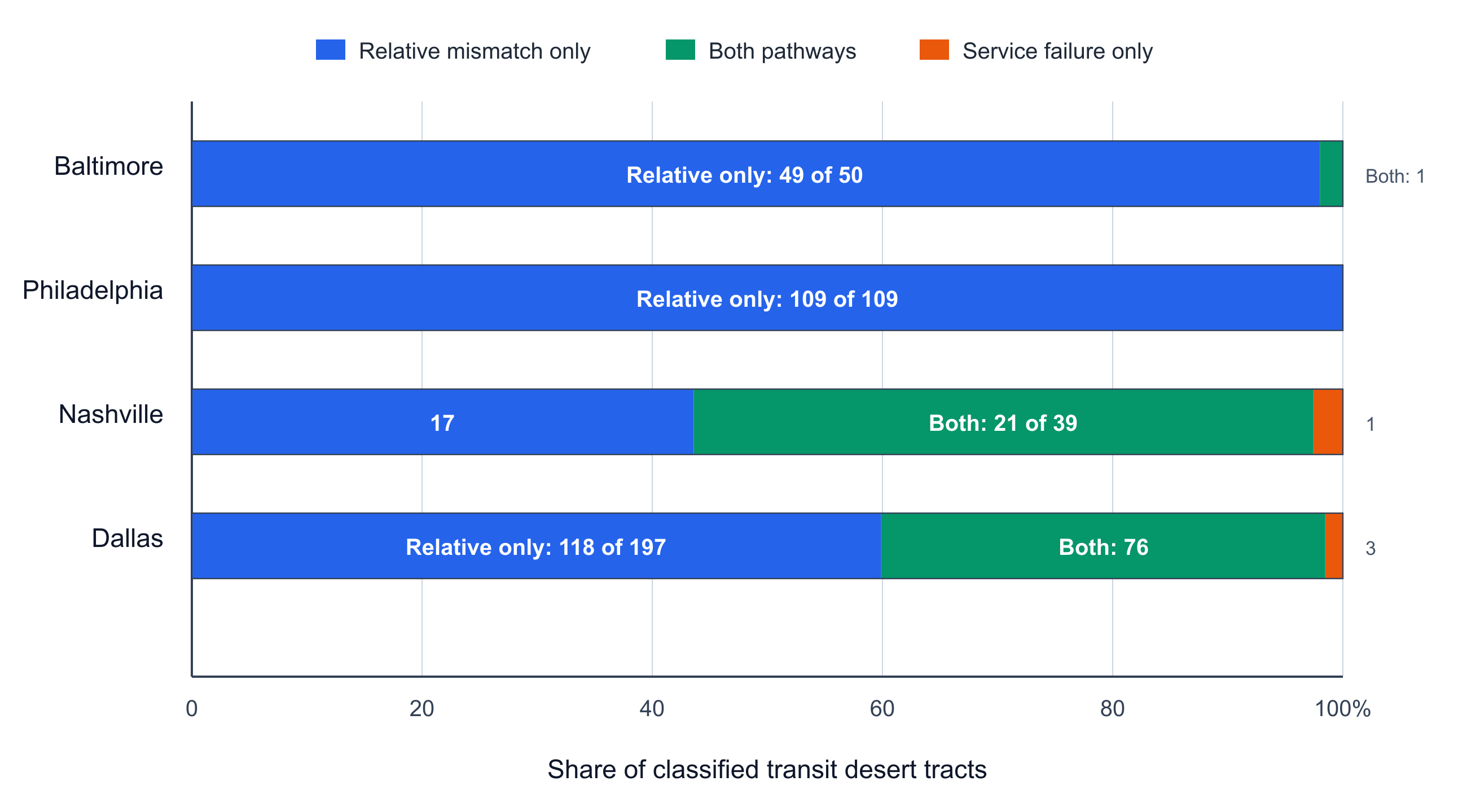}
\caption{Pathway composition among classified transit desert tracts.}
\label{fig:pathway}
\end{figure}

Figure~\ref{fig:pathway} shows the composition of classified transit desert tracts by pathway. The legacy cases are dominated by relative mismatch. In Baltimore, 49 of 50 classified tracts are identified by the relative pathway only, with one by both. In Philadelphia, all 109 classified tracts are relative-only. This suggests classification there is driven primarily by within-city mismatch between supply and measured vulnerability rather than by widespread failure of minimum service. The Sunbelt cases differ. In Nashville, 17 of 39 classified tracts are relative-only, 21 both, and one service-failure-only; in Dallas, 118 of 197 are relative-only, 76 both, and three service-failure-only.  The service-failure pathway rarely acts alone, but it changes the interpretation of many tracts by showing where relative mismatch coincides with failure of minimum service criteria.

\subsection{Minimum-Service Failures}
\begin{figure}[t]
\centering
\includegraphics[width=\columnwidth]{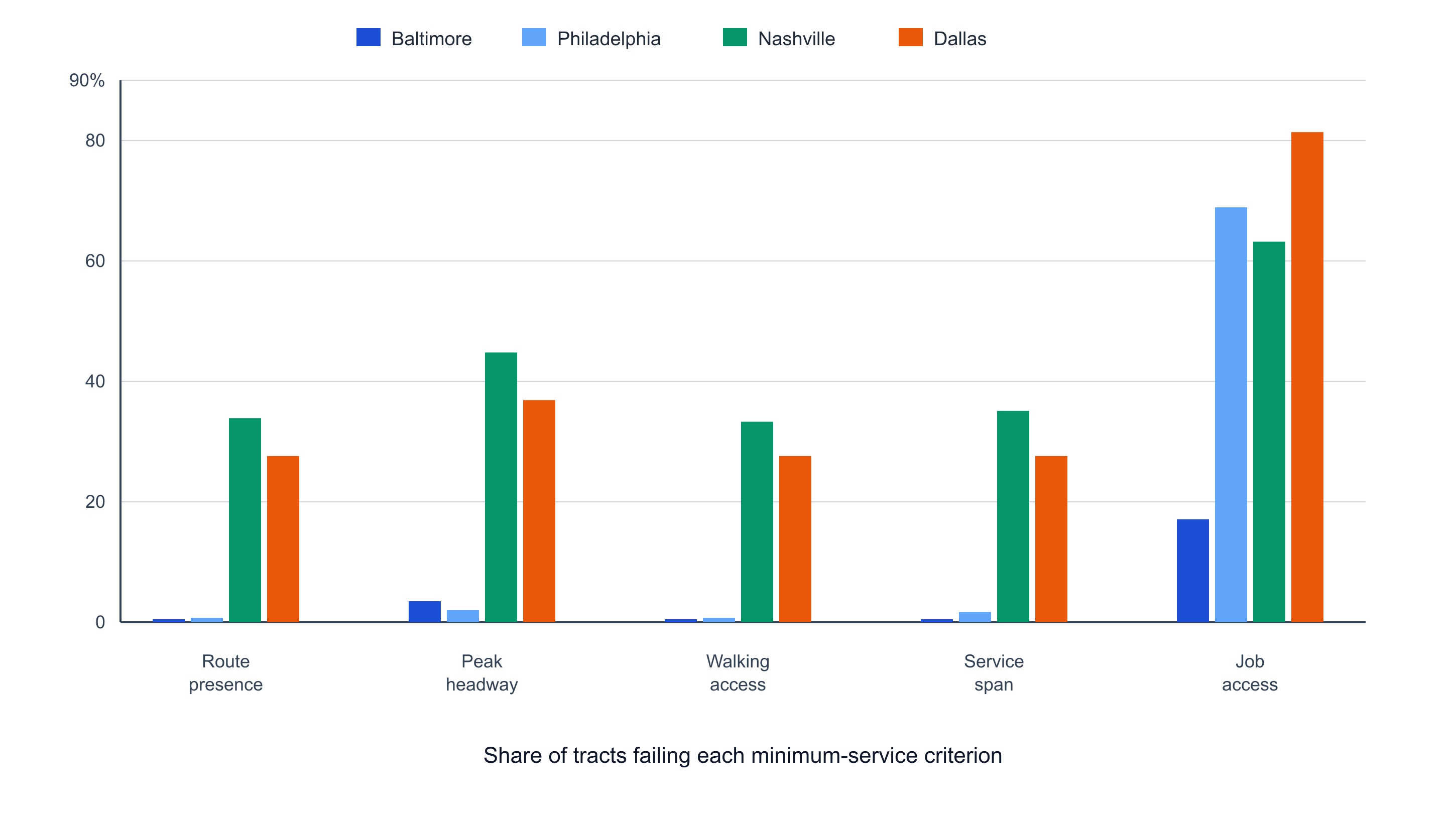}
\caption{Percentage of tracts in each city that fail each 
minimum-service criterion. Nashville and Dallas show higher 
failure rates across most criteria, explaining why the 
service-failure pathway is more active in these cities.}
\label{fig:minservice}
\end{figure}

Figure~\ref{fig:minservice} reports the share of all tracts in each city failing each minimum-service criterion. The strongest contrast is between the legacy and Sunbelt cases. Baltimore has very low failure rates for route presence, walking access, service span, and peak headway, with job accessibility the main exception; Philadelphia similarly has low failure on service-presence measures but a high job-access failure rate. These patterns are consistent with the pathway composition in Fig.~\ref{fig:pathway}: in the legacy cases, desert labels are rarely driven by broad absence of basic service. Nashville and Dallas show substantially higher failure rates across peak headway, service span, route presence, walking access, and job access. A purely relative method can identify which tracts are worse than others within a city, but it does not by itself show how many tracts fall below minimum service conditions, which is what users in these systems actually encounter.

\subsection{User-Experience Deficit Profiles}
\begin{figure}[t]
\centering
\includegraphics[width=\columnwidth]{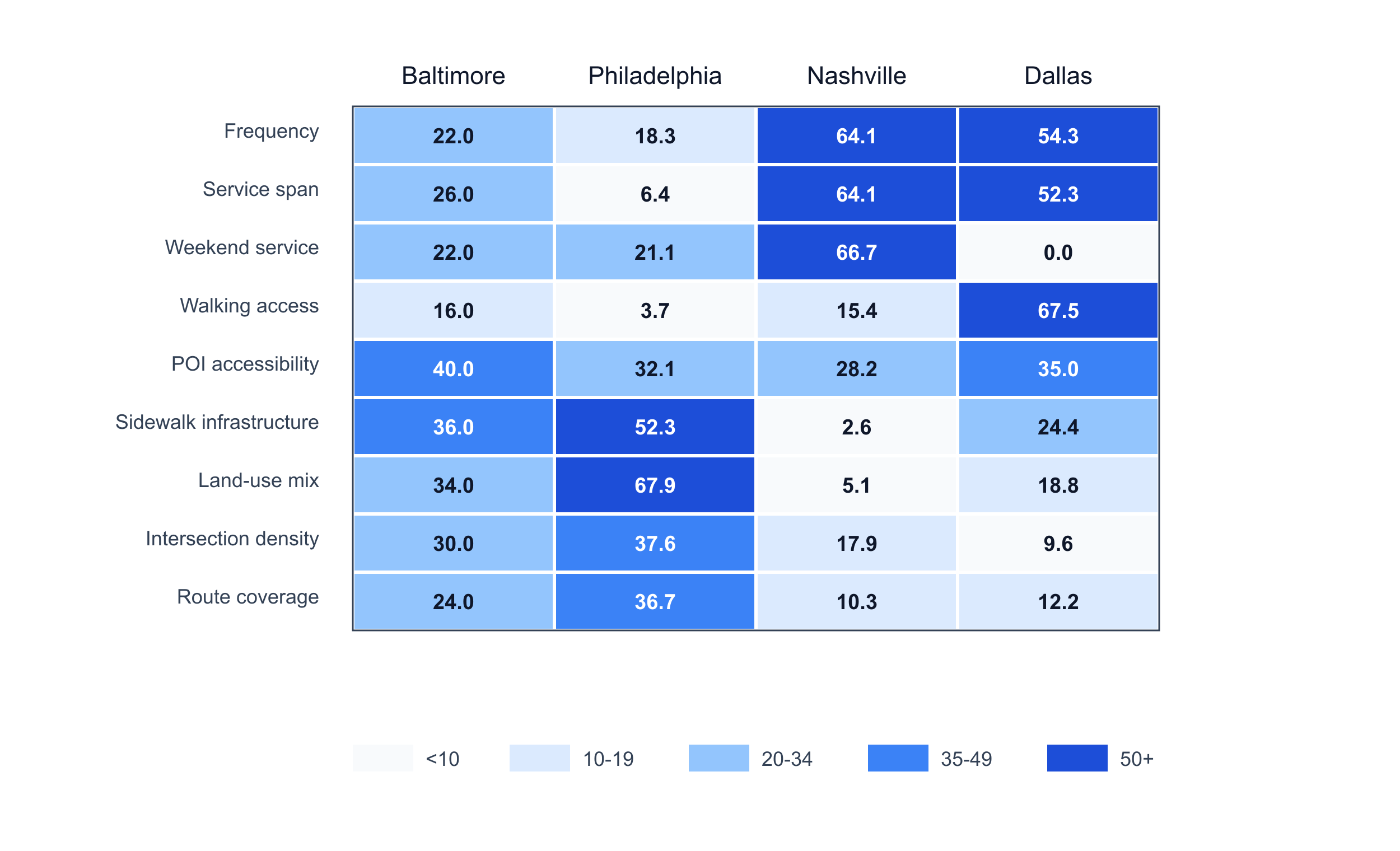}
\caption{Supply-deficit profiles among classified transit desert tracts. Cell values show, for each city, the percentage of that city's classified transit desert tracts in which a feature appears among the three lowest supply scores.}
\label{fig:deficit}
\end{figure}

Figure~\ref{fig:deficit} summarizes supply-deficit profiles among classified tracts; each cell reports the percentage of classified desert tracts in which a feature appears among the three lowest supply scores.  The legacy cases show a built-environment and destination-access signature. In Baltimore, POI accessibility appears among the top deficits for 40.0\% of classified tracts, followed by sidewalk infrastructure, land-use mix, and intersection density; in Philadelphia, land-use mix, sidewalk infrastructure, intersection density, route coverage, and POI accessibility are prominent. Classified tracts here are not places with no transit; users have service but weaker destination access or walkability around it. The Sunbelt cases show a basic-service signature. In Nashville, weekend service, frequency, and service span dominate; in Dallas, walking access, frequency, and service span are most prominent, with POI accessibility also frequent. Where the service-failure pathway plays a larger role, the dominant deficits involve core availability and usability, the dimensions riders feel most directly.

\subsection{Summary of Findings}
Across the four cases, the diagnostic produces three main findings. First, pathway composition differs across transit contexts: Baltimore and Philadelphia are dominated by relative mismatch, while Nashville and Dallas show more overlap between relative mismatch and minimum-service failure. Second, threshold failures show that lower-baseline systems have broader minimum-service deficiencies not captured by relative ranking alone. Third, deficit profiles make classifications more interpretable by showing whether classified tracts are associated primarily with built-environment and destination-access limitations or with basic-service limitations such as frequency, span, weekend service, and walking access. These results are evidence from four contrasting cases, not a universal typology, but they support the central argument that a single under-service label can mask different mechanisms.

\section{Discussion}
The results suggest that under-service labels are more actionable for user-centered mobility when the classification pathway is explicit. In Baltimore and Philadelphia, most classified tracts reflect relative mismatch: service deficits defined primarily in relation to local demand and citywide supply. In Nashville and Dallas, a larger share also fails minimum-service criteria, indicating relative mismatch overlaps with broader baseline service absence. This matters because the same binary label implies different problems for users. A tract identified only through relative mismatch may need service realignment, better pedestrian connection to existing service, or improved destination access. A tract identified through both pathways may indicate a more fundamental lack of adequate transit for vulnerable users, pointing toward new coverage, frequency, and span.

For Mobility-as-a-Service platforms and demand-responsive service 
designers, the pathway distinction has practical implications. 
Relative-mismatch tracts, where some service exists but is 
misaligned with need, may benefit from demand-responsive overlays, 
improved real-time information, or multimodal integration connecting 
existing routes to underserved demand. Service-failure tracts, where basic coverage is absent, need new service investment -- whether fixed-route expansion, 
microtransit, or demand-responsive options. 

The deficit profiles add further specificity by translating labels into user-facing service attributes and concrete intervention directions. Built-environment and destination-access deficits (prominent in the legacy cases) point toward pedestrian infrastructure, land-use coordination, or network-connectivity improvements, consistent with evidence that street design and traffic stress shape whether users choose to walk or ride~\cite{bas2023}; basic-service deficits (prominent in the Sunbelt cases) point more directly toward coverage, frequency, and span investments. In this sense the diagnostic can act as a decision-support layer for user-centered smart-mobility planning, informing service-zone targeting~\cite{erdogan2024}, Mobility-as-a-Service gap analysis, equity-aware governance, and citizen engagement rather than acting as a black-box label. Because it runs entirely on open data, it is transferable across cities and repeatable as service and demographics change.

The workflow has several limitations. It is a screening tool, not a causal model: it identifies where different forms of under-service appear but does not estimate intervention effects. Results depend on public-data quality, including GTFS completeness, ACS sampling uncertainty, LEHD employment data, and OpenStreetMap coverage. Because the workflow relies on GTFS static schedules, it captures scheduled service availability rather than realized reliability; future extensions should incorporate GTFS-RT, on-time performance, missed trips, cancellations, and crowding where available. As a robustness check, we varied the minimum-failure rule (two, three, or four of five criteria), the peak-headway threshold (20, 30,  or 45 minutes), and the job-access threshold (2{,}500, 5{,}000, or 10{,}000 reachable jobs). Both-pathways classification rates were unchanged in Baltimore and Philadelphia and shifted only modestly in Nashville and Dallas, from 22.4--23.6\% and 30.5--31.8\%, respectively. This indicates that the main pathway interpretation is not driven by a single cutoff: the legacy cases remain dominated by relative mismatch, while the lower-baseline cases retain greater overlap between relative mismatch and minimum-service failure. Even so, the threshold values in the service-failure pathway are reasonable screening benchmarks but should be validated with agencies, communities, and local planning standards before use in formal prioritization. Future validation could compare pathway labels and deficit profiles with rider surveys, complaint logs, stop-level ridership, agency service standards, and practitioner review, to check whether the diagnosed deficits align with lived experience and local planning priorities. Finally, additional cities and longitudinal data are needed to test whether the pathway distinctions hold across other contexts.

\section{Conclusion}
This paper frames transit under-service detection as a user-centered diagnostic problem for inclusive smart mobility. Relative and absolute approaches capture different aspects of the rider's experience: relative methods identify local mismatch, while absolute thresholds reveal minimum-service failure. We combine these perspectives in a reproducible open-data workflow that constructs supply and user-need features, identifies spatially clustered mismatch, applies an equity-informed service-failure pathway, and produces pathway labels with user-relevant supply-deficit profiles. Applied to Baltimore, Philadelphia, Nashville, and Dallas, the diagnostic shows that the same under-service label can reflect different mechanisms: the legacy cases are identified primarily through relative mismatch, while the Sunbelt cases show greater overlap with minimum-service failure, and deficit profiles distinguish built-environment and destination-access limitations from basic-service limitations. These findings suggest that transit desert maps should report not only \emph{where} under-served areas are, but \emph{why} they are classified, in terms of the service attributes users actually experience. Pathway-based classification does not replace local planning judgment, formal equity analysis, or community engagement, but it provides a more interpretable, user-centered screening output for comparing mobility gaps across cities with different transit baselines. Future work should test the diagnostic in additional cities, incorporate service reliability, and validate deficit profiles 
against rider surveys and lived experience.

\section*{Code and Data Availability}
The workflow uses publicly available data sources, including GTFS feeds, ACS 5-Year Estimates, TIGER/Line tract boundaries, LEHD LODES employment data, and OpenStreetMap-derived features. Code, configuration files, and processed outputs needed to reproduce the analyses are available at \url{https://github.com/Davidavid45/transit-desert-pipeline-extended}.

\section*{Acknowledgment}

This work was supported in part by the U.S. National Science Foundation under Grant 1951924.

\bibliographystyle{IEEEtran}
\bibliography{refs}

\end{document}